\documentstyle[aps,epsfig,floats]{revtex}
\begin{document}
 
\title{On modulational instability and energy localization \\
in anharmonic lattices at finite energy density}

\vspace{0.5cm}

\author{Yuriy A. Kosevich\thanks{On leave from: Moscow State Technological
University ``STANKIN'', 101472 Moscow, Russia; 
E-mail: {\tt yukos@mpipks-dresden.mpg.de}}}
\address{Max Planck Institute for Physics of Complex Systems\\ 
Noethnitzer Str.38, D-01187 Dresden, Germany}

\author{Stefano Lepri \thanks{Correspondig author; also with 
Istituto Nazionale di Fisica della Materia, Unit\`a di Firenze; 
E-mail: {\tt lepri@avanzi.de.unifi.it}}}

\address{Dipartimento di Energetica ``S. Stecco''
Via Santa Marta 3, I-50139 Florence, Italy}

\date{\today}
\maketitle 
\begin{abstract}
The localization of vibrational energy, induced by the modulational instability of
the Brillouin-zone-boundary mode in a chain of classical anharmonic 
oscillators with finite initial energy density,  is studied 
within a continuum theory. We describe the  initial localization stage as 
a gas of envelope solitons and explain their merging, eventually leading
to a single localized object containing a macroscopic fraction of 
the total energy of the lattice.  The initial-energy-density dependences of all 
characteristic time scales of the soliton formation and merging are described
analytically.  Spatial power spectra are computed and used for the
quantitative explanation of the numerical results.

\vspace{0.2cm}
\noindent{\sf Submitted to Phys. Rev. B}\\
\end{abstract}
\vspace{0.2cm}

\section{Introduction}

Despite of a large number of studies on localization of energy in  
pure anharmonic lattices \cite{KosKov,Dolg,SivPage,Flach}, the mechanism of their thermal
generation is still under debate. This is clearly a central issue 
being directly related with the possibility of observing signature 
of such a class of excitations in experiments on real crystals. The problem of
nanoscale energy localization has turned out to be rather important also in view
of its possible application for different physical and biological systems. 
The theoretical studies at finite energy per particle  
dealt mainly with one-dimensional models
belonging both to the Fermi-Pasta-Ulam (FPU) class of model 
\cite{Burlakov,Stefano} (with acoustic spectrum in the harmonic limit) and to 
the Klein-Gordon one \cite{Peyrard} (with optical-phonon-type spectrum in the 
harmonic limit). Some attention has been 
also recently payed to the discretized version of the nonlinear Schr\"odinger 
equation in one dimension \cite{Bishop,Rasmu}.

Of special interest is the numerically observed fact that excitation of 
short-wavelength
extended modes of the chain and their instability with respect to perturbations
with nonvanishing wavenumber leads generically to creation of localized states.
The phenomenology emerging from the numerical studies 
\cite{Burlakov,Stefano} can be summarized in three main stages.
In the first one, after the development of the modulational instability, 
energy localizes along the chain in the form of an array of almost
standing ``bumps''. After a given time (that increase upon
decreasing the temperature) those begin to merge diminishing
their number and localization lengths. This second stage (which is  
reminiscent of the phenomenon of  ``soliton turbulence'' \cite{Kingsep})
ends with the emergence of a single localized and moving object (a 
``breather''), carrying along a macroscopic fraction of the total 
energy. In the third stage, the breather continues to slow down
radiating its energy towards the background until 
it eventually disappears and the 
equipartition (thermal equilibrium) state  is attained.

The aim of this paper is to give a consistent quantitative explanation of such behaviour, 
in particular of the apparent paradoxical tendence of the system 
to increase localization as time goes by. Namely, we would like to 
clarify the role of the finite vibrational energy density on the formation of
strongly localized envelope solitons and breathers and their temporal
evolution in the anharmonic FPU chain. This is accomplished by
means of a suitable continuum approximation for the lattice dynamics.
The approach in terms of the envelope-function  equation including higher-order
nonlinear terms has been already introduced
to describe short-wavelength excitations in the FPU chain \cite{KosPLA} and is reviewed
in Section II. It is a useful framework both for computing 
localized solution in the form of envelope soliton (ES) and breather 
(section III), and for describing the modulational instability 
(Section IV). Furthermore, going beyond the usual integrable approximation 
\cite{Yoshi} is of 
crucial importance to explain the localization process (Sections V and VI). We
described
analytically the initial-energy-density dependence of all 
characteristic time scales of the envelope soliton formation and merging. 
Particular emphasis has been also given to the analytic computation and
numerical measurements of 
the spatial power spectra (``structure functions'') that, at least in principle, 
are of importance for real experiments. Indeed, they turn out to be  
extremely useful for the  physical interpretation of the simulation results 
(see, e.g.,  Sections V and VII). Finally, a brief discussion 
of the finite-temperature effects is reported in Section VIII.

\section{Envelope-function equation for the FPU model}

We start from the classical FPU 
Hamiltonian of a monatomic chain of $N$ particles of mass $m$ 
with anharmonic nearest-neighbor potential of order $r\geq 3$:
\begin{equation}
{\cal H}=\sum_{n}\left[\frac{1}{2}m \dot u_n^2 +\sum_{\gamma =2}^{r}
\frac{K_{\gamma}}{\gamma} (u_{n+1}-u_{n})^{\gamma}\right]~,
\label{hami}
\end{equation}
where $u_{n}$ is the (real scalar) displacement of the $n-$th particle 
from its  equilibrium position and $K_{\gamma}$ are the harmonic
($\gamma =2$) and anharmonic force constants. The lowest order potential
describing the anharmonicity of the longitudinal or pure transverse motion in
a centrosymmetric $1D$ lattice corresponds, respectively, 
to $r=3$ or $r=4$. From the Hamiltonian (\ref{hami}) we obtain the following
equations of motion:
\begin{equation}
m\ddot{u}_{n}=\sum_{\gamma =2}^{r}
K_{\gamma}[(u_{n+1}-u_{n})^{\gamma -1}-(u_{n}-u_{n-1})
^{\gamma -1}]~.
\label{eqmot}
\end{equation}  
In the following, periodic boundary conditions are always assumed.

In order to describe the  short-wavelength excitations of the chain
with wavenumber $k\sim \pi/a$ ($a$ being the lattice spacing), i.e. near the 
Brillouin-zone boundary, it is convenient to introduce an
envelope-function of the displacements $f_{n}=(-1)^{n}u_{n}$, 
$f_{n}\equiv f(x/a)$. The latter is assumed to be slowly varying  on the 
interatomic scale: $a f_x\ll f$. Substituting in Eq. (\ref{eqmot}) the
expansion of the differences $u_{n\pm 1}-u_{n}$ up to the fourth order, one 
can obtain a nonlinear partial differential 
equation for $f$ $\cite{KosPLA}$. For instance, in the case of the 
purely quartic anharmonicity (i.e. the so-called $\beta$-FPU model with 
only $K_2$ and $K_4$ different from zero) such an equation reads 
\begin{equation}
m\ddot{f}+4K_{2}f+K_{2}a^2 f_{xx}+ 
\frac{1}{12}
K_{2}a^4 f_{xxxx} +16K_{4}f^{3}+6K_{4}a^{2}f(f^{2})_{xx}=0~,
\label{envelope}
\end{equation}   
where the subscript $x$ stands for the spatial derivative of the corresponding
order. Equation, similar to Eq. (\ref{envelope}),  
can be also obtained for the cubic anharmonicity  ($\alpha$-FPU model
with $\gamma=2$ and $\gamma=3$) and, more generally,  for the FPU chain 
(\ref{hami}) 
with the anharmonic potential of the arbitrary (even or odd) order $r$ 
$\cite{KosPRB}$. 

Due to the last term in l.h.s. of Eq. (\ref{envelope}), the latter substantially 
differs from the conventional nonlinear Klein-Gordon and its nonrelativistic 
version - the nonlinear Schr\"odinger equation. Actually, Eq. (\ref{envelope}) can be
reduced to the latter only in the limit of relatively small oscillation amplitudes 
(see Section $IV$ below). Eq. (\ref{envelope}) differs also from the well-known
Korteweg-de Vries equation which describes the evolution 
of the long-wavelength (i.e. the ``acoustic-phonon-type") excitations and can 
be derived from the lattice Hamiltonian (\ref{hami}) in the limit $au_{x}\ll 1$ 
(see, e.g., \cite{KosPRB}).   

In what follows, we will use Eq. (\ref{envelope}) for the adequate $analytic$ 
description of all the main stages of 
modulation-instability-induced dynamic evolution of the hard (with $K_{4}>0$) 
quartic FPU chain. For simplicity of notation, we set $K_2$, $K_4$, $a$, and $m$ 
to unity. With such a choice, the (adimensional) energy density (i.e. the energy 
per particle) $\epsilon$ is the only relevant physical parameter if one does
not account for the finite-temperature effects (see Section $VIII$ below). 
Moreover, we will always deal with the (classical) low-temperature limit that in our 
units corresponds to $\epsilon\ll 1$.

\section{Localized solutions: envelope solitons and breathers}

As a first step, we determine the form of the 
single ES (or periodic array of solitons) by looking for solutions of 
Eq. (\ref{envelope}) in the monochromatic form $f(x,t)=\phi(x)\cos\omega t$.
The frequency $\omega$ can be computed by first substituting such an Ansatz 
in (\ref{envelope}) and invoking then the so-called Rotating Wave Approximation 
(RWA), namely to replace $\cos^3\omega t$ by $\frac{3}{4}\cos\omega t$.
It is worth mentioning 
that such an approximation is expected to hold for the considered 
short-wavelength vibrations due to the weakness of nonresonant  
interaction between the mode with fundamental frequency and its third harmonic. 
By further neglecting the fourth-order dispersive term in Eq. (\ref{envelope}) 
we find its first integral as $\cite{KosPLA}$:
\begin{equation}
(4-\omega^2)\phi^2+(1+9\phi^2)(\phi_{x})^2+6\phi^4=const~.
\label{intmot}
\end{equation}
The generic solution of Eq. (\ref{intmot}) for a nonvanishing constant corresponds to a periodic array of ESs, and can be conveniently parametrized by the 
two extreme values of the amplitude $\phi_{min}$ and  $\phi_{max}$   
(so that $\phi_{min}\leq\phi\leq\phi_{max}$). Those are simply determined by the  
conditions that $\phi_{x}$ vanishes for $\phi=\phi_{min}$ and 
$\phi=\phi_{max}$. From those two conditions we derive immediately the 
frequency of the envelope-soliton array as a function of the extreme values of
its amplitude: 
\begin{equation}
\omega^2=4+6(\phi_{min}^2+\phi_{max}^2)~.
\label{omegas}
\end{equation} 
Then from Eqs. (\ref{intmot}-\ref{omegas}) we can determine the desired
solution as 
\begin{equation}
\left(\frac{d\phi}{dx}\right)^2=6\frac{(\phi_{max}^2-\phi^2)
(\phi^2-\phi_{min}^2)}{1+9\phi^2}~. 
\label{intsol}
\end{equation}
Eq. (\ref{intsol}) can be solved by quadratures yielding the following 
implicit form for $\phi(x)$:
\begin{equation}
|x-x_0|\;=\;\frac{1}{\sqrt{6}}\int_{\phi}^{\phi_{max}}
\frac{\sqrt{1+9\phi^{'2}}}{\sqrt{(\phi_{max}^2-\phi^{'2})
(\phi^{'2}-\phi_{min}^2)}}d\phi^{'}~, \qquad\phi(x_0)=\phi_{max}~.
\label{implicit}
\end{equation} 
The single ES corresponds to $const=0$ in Eq. (\ref{intmot})  
and to $\phi_{min}=0$ in Eq. (\ref{implicit}). Although a  general 
explicit form of Eq. (\ref{implicit}) is not feasible, we can at least 
discuss the two limiting cases of small and large amplitudes.  
In the small-amplitude limit ($9\phi_{max}^2\ll 1$), the solution is 
given by (we include explicitely the time dependence)
\begin{equation}
f_s(x,t)\;=\;\phi_{max}\frac{\cos(\omega_{s} t+\varphi)}
{\cosh[(x-x_0)\phi_{max}\sqrt{6}]}~,
\label{soliton}
\end{equation}
where the frequency as given by Eq. (\ref{omegas}) with $\phi_{min}=0$
and is approximately $\omega_s=2+3\phi_{max}^2/2$, whereas $\varphi$ is 
the initial phase. 

In the large-amplitude limit ($9\phi_{max}^2\gg 1$), it follows 
from Eq. (\ref{implicit}) that the ES acquires  a sinusoidal shape with a short 
amplitude-independent localization length \cite{KosPLA}
\begin{equation}
f_b(x,t) \;=\; \phi_{max}\cos(\omega_{b} t+\varphi)\cos[k_{b}(x-x_0)] \quad
{\rm for} \quad  
k_{b}|x-x_{0}|<\frac{\pi}{2}
\label{breather}
\end{equation}
and with $f_b\approx 0$ outside of the above specified domain. The effective 
wavenumber is $k_{b}=\sqrt{2/3}$ and $\omega_{b}$ given by Eq. (\ref{omegas}) 
with $\phi_{min}=0$. This is rather accurate approximation for  
the exact breather solution of the discrete FPU model (see Ref. \cite{KosPRB}
for a detailed comparison). Discrepancies have to be expected because, 
strictly speaking, the continuum approach is not completely 
justified in this 
short-wavelength limit. Indeed, the analysis of the corresponding 
discrete-lattice 
equations predicts \cite{KosPRL}  the existence of intrinsic localized modes
(discrete breathers) with 
(approximately) sinusoidal envelope (\ref{breather}) but with $k_b=\pi/3$ and 
$\omega_{b}^2=3+(81/16)\phi_{max}^2$.
 
Having at hand the explicit solution (\ref{soliton}), we can 
compute the spatial Fourier transform of the displacement field $u$ associated 
to the single ES obtaining 
\footnote{Notice that, by definition, the zero wavenumber for $f$ 
corresponds to the wavenumbers $\pm\pi$ for $u$ and therefore 
$
u(k)=\int u(x)e^{ikx}dx=\int f(x)e^{ikx}\cos\pi x dx = 
\frac{1}{2} [f(k-\pi)+f(k+\pi)]\quad.
$
}
\begin{equation}
u_{s}(k,t)\;=\;\frac{1}{2}\cos(\omega_{s} t +\varphi)
\left[F_s(k-\pi,x_0) + F_s(k+\pi,x_0) \right] \quad,
\label{spesol}
\end{equation}
where we have defined the form factor
\begin{equation}
F_s(k,x_0)\;=\;\frac{\pi}{\sqrt{6}}
\frac{e^{ikx_0}}{\cosh[\frac{\pi k}{\phi_{max}2\sqrt{6}}]} \quad.
\label{spesol2}
\end{equation}
Accordingly, the power spectrum (i.e. the square modulus of $u_{s}(k,t)$) is  
exponentially localized around $k=\pm\pi$ and its maximal value is independent 
of the amplitude $\phi_{max}$.

For later purposes, we compute also the total energy $E_s$ of the ES. To 
this aim, one can use the expression for the hamiltonian density $h$ by the 
variation of which the equation (\ref{envelope}) can be derived \cite{KosPLA} 
\begin{equation}
h=\frac{1}{2}(\dot{f})^2+2f^2+
4f^{4}-\frac{1}{2}(1+12f^2)
(f_{x})^2+\frac{1}{24}
(f_{xx})^2, 
\label{hamidens}
\end{equation}
(remember that we are working in the dimensionless units). Neglecting again 
the higher-order dispersive term and averaging $h$ over time (in the RWA) 
and by means of (\ref{intmot}) with $const=0$, we find that the energy density
$h_{s}$ 
associated with the ES is $2\phi^2+3\phi^4$. Using this 
expression together with (\ref{soliton}), we finally find
\begin{equation}
E_{s}\;=\;\int_{-\infty}^{+\infty}\, h_{s}dx=\;\int_{-\infty}^{+\infty}\, \left[ 2\phi^2+3\phi^4 \right] \, dx
\;=\;\frac{4}{\sqrt{6}}\, \phi_{max}(1+\phi_{max}^2). 
\label{es2}
\end{equation}  
In the leading approximation, the energy scales linearly with the amplitude   
$\cite{Kingsep}$. 
 
The Fourier transform of the breather solution (\ref{breather}) is
\begin{equation}
u_{b}(k,t)\;=\;\frac{1}{2}\cos(\omega_{b} t +\varphi)
\left[F_b(k-\pi,x_0) + F_b(k+\pi,x_0) \right] \quad,
\label{spebrea}
\end{equation}
where we have defined
\begin{equation}
F_b(k,x_0)\;=\;\phi_{max}\, 
e^{ikx_0}\, \frac{2k_b}{k_b^2-k^2}\, \cos\left(\frac{\pi k}{2k_b}\right) \quad.
\end{equation}
Notice that this function is finite for $k=k_b$. In particular, for
$k_b=\pi/3$ the Fourier transform of the breather solution has the following
characteristic form:
\begin{equation}
u_{b}(k,t)\;=\frac{\pi}{3}\phi_{max}\cos(\omega_{b} t
+\varphi)\sin(\frac{3}{2}k)
\,e^{ikx_0}\, 
\left[\frac{e^{-i\pi x_0}}{(k-4\pi/3)(k-2\pi/3)}-\frac{e^{i\pi
x_0}}{(k+4\pi/3)(k+2\pi/3)}\right]~. 
\label{spebreaex}
\end{equation}

The power spectrum has basically
a sinusoidal shape with two absolute maxima at $k=\pm \pi$ whose height is proportional 
to the square of the breather amplitude. 

\section{Modulational instability}

In this section, we describe the modulational instability of the
Brillouin-zone-boundary mode corresponding to the uniform envelope 
$f(x,t)\equiv f_0(t)=A\cos\omega t$. To compute the frequency $\omega$ one can resort 
again to the RWA mentioned in the previous section. In the small-amplitude 
limit $3A^2\ll 1$ one obtains $\omega=2+3A^2$, which amounts to a correction to the 
zone-boundary frequency $\omega_{max}=2\sqrt{K_{2}/m}\equiv 2$. 

Let us consider a spatially inhomogeneous small perturbation $\delta f\propto\cos k_{m}x$, 
where the wavenumber $k_{m}$ of the envelope-function modulation is 
assumed to be much smaller than the Brillouin-zone width $\pi$, an hypotesis that 
will be confirmed {\it a posteriori}. Then, from the linearization of 
Eq. (\ref{envelope}) we find the equation for $\delta f$ to be
\begin{equation}
\delta\ddot{f}+[4-k_{m}^2+24A^2(1+\cos 2\omega t)]\delta f\;=\;0~.
\label{paramet}
\end{equation} 
To obtain the growth rate $s$ at the parametric resonance, described by 
Eq. (\ref{paramet}), we take a solution of this equation as (cf., e.g., 
$\cite{L.L.Mech}$), 
\begin{equation}
\delta f=[a\cos \omega t + b\sin \omega t]
e^{st}\cos k_{m}x~, 
\label{growth} 
\end{equation}
where the unknown amplitudes $a$ and $b$  are assumed to be small with respect to 
$A$. Therefore, from Eqs. (\ref{paramet}) and (\ref{growth}) 
we obtain the  dispersion equation for the growth rate $s$:
\begin{equation}
s^4+2s^2[8+36A^2-k_{m}^2]=k_{m}^2(24A^2-k_{m}^2)~.
\end{equation}
 
The maximal growth rate $s_{max}$ 
and the corresponding wavenumber $k_{m}^{\star}$ of the envelope-function 
modulational instability are thus given in the small-amplitude limit $3A^2\ll 1$ by  
\begin{equation}
k_{m}^{\star}=A\sqrt{12}~, ~~~ s_{max}=3A^2~,~~~ b=-a~, 
\label{kstar}
\end{equation} 
which, consistently, describes a longwavelength modulation with $k_{m}^{\star}/2\ll 1$
\cite{Sand.Page2}. In the considered limit, such a value of 
$s_{max}$ coincides with the one 
computed from the stability analysis of the discrete-lattice equations
\cite{Stefano2}. It is also worth to mention that 
there is a minimal (threshold) amplitude $A_{c}$ for the
modulational instability  of the
Brillouin-zone-boundary mode in a finite FPU chain of $N$ particles, and 
$A_{c}\propto 1/N$ for $N\gg 1$, see, e.g. \cite{Stefano}. 
We neglect such finite-size effects 
in our continuum  approach 
which implies, in particular, that the amplitude $A$ in Eq. 
(\ref{kstar})  
is well above the threshold one for the considered (finite) chain. 

By means of Eq. (\ref{envelope}) we can also describe the interesting phenomenon 
of the further appearance of spatial harmonics of the ``primary'' perturbation 
i.e. the components with wavenumbers $nk_{m}^{\star}$ and $n=0,2,3,...$. To
describe for instance the growth of the zeroth and second spatial harmonics, 
we assume $\delta f$ to be of the following form (cf. Eqs. (\ref{growth}) 
and (\ref{kstar})):
\begin{eqnarray}
\delta f&=&a\left[\cos \omega t -\sin \omega t \right]
e^{s_{max}t}\cos k_{m}^{\star}x + 
\left[c\cos \omega t + d\sin \omega t \right]
e^{2s_{max}t}\cos 2k_{m}^{\star}x \nonumber \\
&&+\left[e\cos \omega t  + g\sin \omega t \right]e^{2s_{max}t}
\equiv\delta f_{k_{m}^{\star}}+\delta f_{2k_{m}^{\star}}+
\delta f_{0}~,
\label{deltaf2}
\end{eqnarray}
where the amplitudes $c$, $d$, $e$, and $g$  are assumed to be much smaller 
than both the condensate amplitude $A$ and the amplitude $a$ of the
primary perturbation. Using Eq. (\ref{envelope}), 
we obtain the following evolution equations for 
$\delta f_{2k_{m}^{\star}}$ and $\delta f_{0}$ similar to 
Eq. (\ref{paramet}) with a driving term $-48f_{0}(\delta f_{k_{m}^{\star}})^2$ on 
the r.h.s.:
\begin{eqnarray}
&&\delta\ddot{f}_{2k_{m}^{\star}}+[4-4k_{m}^{\star 2}+24A^2(1+\cos 2\omega t) ]
\delta f_{2k_{m}^{\star}} = \nonumber \\
&&-12a^2A[2\cos \omega t 
-\sin \omega t ]e^{2s_{max}t}\cos 2k_{m}^{\star}x ~, \\
&&\delta\ddot{f}_{0}+[4+24A^2(1+\cos 2\omega t )]\delta 
f_{0}=
-12a^2A[2\cos \omega t -\sin \omega t ]e^{2s_{max}t}~.
\label{second}
\end{eqnarray}            
>From Eqs. (\ref{deltaf2})-(\ref{second}) we find the amplitudes $c$, $d$, $e$, and $g$:
\begin{equation}
c=\frac{a^2}{2A}~,  \qquad d=e=-\frac{a^2}{2A}~, \qquad
g=-\frac{3a^2}{4A}~.
\end{equation}
Therefore the zeroth $\delta f_{0}$ and second 
$\delta f_{2k_{m}^{\star}}$ spatial harmonics grow exponentially with a
rate which is two times larger than that of the primary instability  
$\delta f_{k_{m}^{\star}}$. Nevertheless, as their initial amplitudes are of 
second order with respect to the amplitude $a$ of the primary instability, 
one expects that they will come into play only at a later stage. 

More generally, we can show that the growth rate of the 
$n$-th spatial harmonic $\delta f_{nk_{m}^{\star}}$ of the primary instability is 
equal to $ns_{max}$ while its initial amplitude is proportional to the $n$-th power 
of the amplitude $a$. Furthermore, it can be shown that small contributions to the 
growth of the zeroth harmonic $\delta f_{0}$ are induced by the higher-order harmonics. 
These arise from the terms like $(\delta f_{2k_{m}^{\star}})^2f_{0}$, 
$(\delta f_{k_{m}^{\star}})^2\delta f_{2k_{m}^{\star}}$ etc. 
to the r.h.s of Eq. (\ref{second}).    

\begin{figure} 
\centering\epsfig{figure=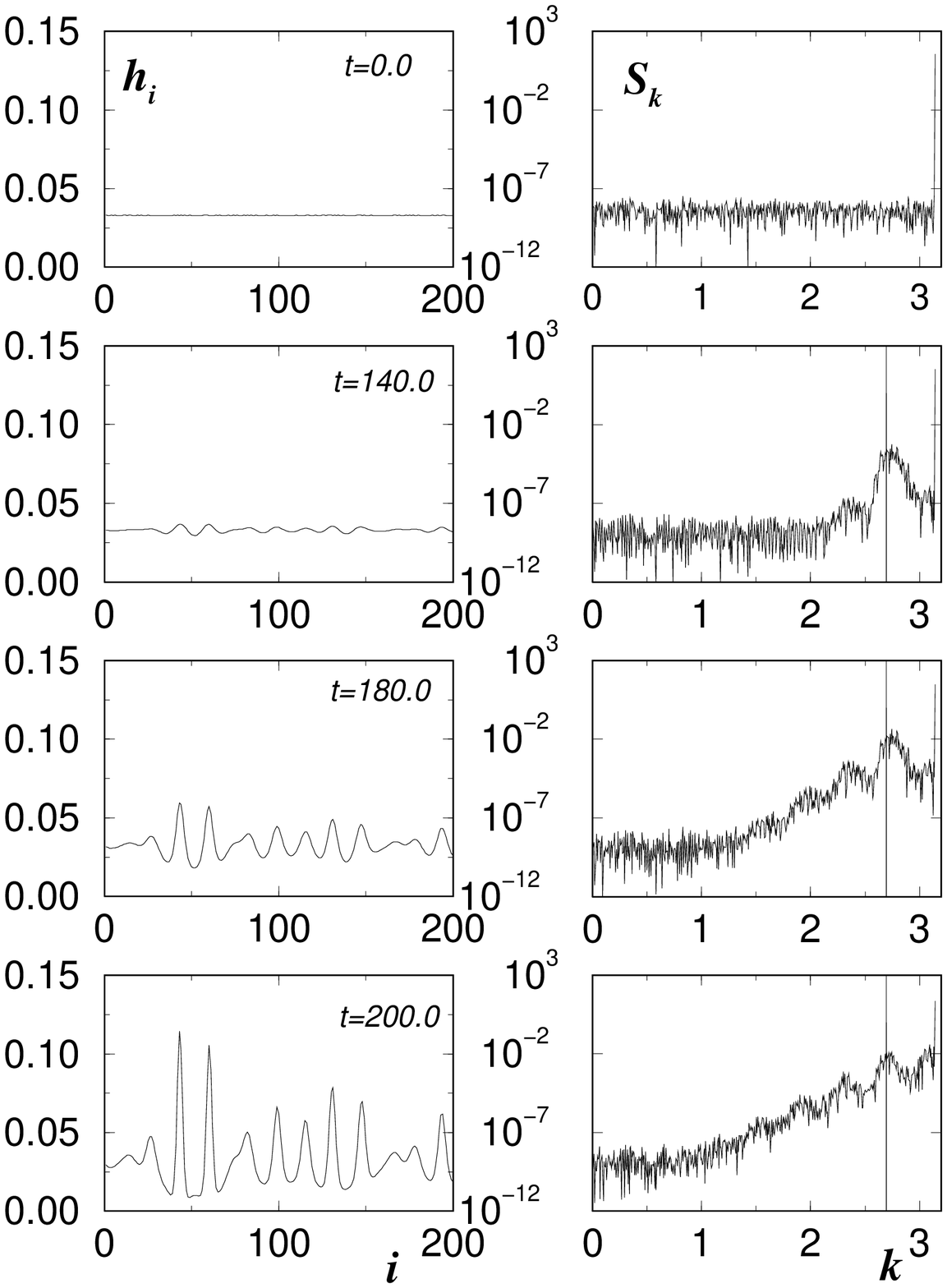,width=12cm}
\caption{
Modulational instability of the boundary mode for 
the quartic FPU model with $\varepsilon=0.0330$, $N=1024$ 
The left panels are snapshots of the energy density, 
the right ones the spatial Fourier spectrum of the velocities. 
The vertical lines correspond to the theoretical value 
of $k_m^*$ given by Eq. (\protect\ref{kstar}). 
}
\label{instabil}
\end{figure}

The above scenario is illustrated in Fig. \ref{instabil}, where the outcomes of 
a numerical simulation of the quartic FPU model are reported. 
The equations of motion were integrated with a third-order symplectic algorithm 
\cite{algor} (microcanonical simulation) starting from the initial conditions  
\begin{equation}
u_n(0)=0, \qquad \dot u_n(0)= (-1)^n\,\sqrt{2\epsilon}~. \qquad
\label{ic}
\end{equation}
Actually, a small gaussian white noise (with amplitude $\ll \sqrt{2\epsilon}$) 
has also been added to the initial velocities in order to speed up the instability.
The left panels of Fig. \ref{instabil} show the evolution of the energy
density (cf. Eq. (1)), 
\begin{equation}
h_n = \frac{1}{2}\dot u_n^2 +\frac{1}{2}\sum_{\gamma =2,4}
\frac{1}{\gamma}\left[(u_{n+1}-u_{n})^{\gamma}+(u_{n}-u_{n-1})^{\gamma}
\right]~ , 
\end{equation}
at subsequent times. The spatial power spectrum of the velocities 
\begin{equation}
S_k \;=\; 
{1\over N}\big| \sum_{n=1}^N \dot u_n \, e^{ikn} \big|^2
\quad, \qquad k=0, \pm {2\pi\over N}, \pm {4\pi\over N}\ldots, \pm\pi
\label{sk}
\end{equation}
shows the growth of the perturbation with the wavenumbers $\pi-k_{m}^{\star}$ 
(vertical line). Obviously a symmetric peak at $-\pi+k_{m}^{\star}$ is also
present 
as well as the further transfer of the energy towards the  spatial harmonics
with wavenumbers $nk_{m}^{\star}$, $n=0,2,3,...$, 
according to the above described mechanism. We also checked that the typical 
time for developing the instability scales as 
$A^{-2}\propto \epsilon^{-1}$ as prescribed by Eq. (\ref{kstar}). 

\section{The first localization stage: a soliton gas}

\begin{figure} 
\centering\epsfig{figure=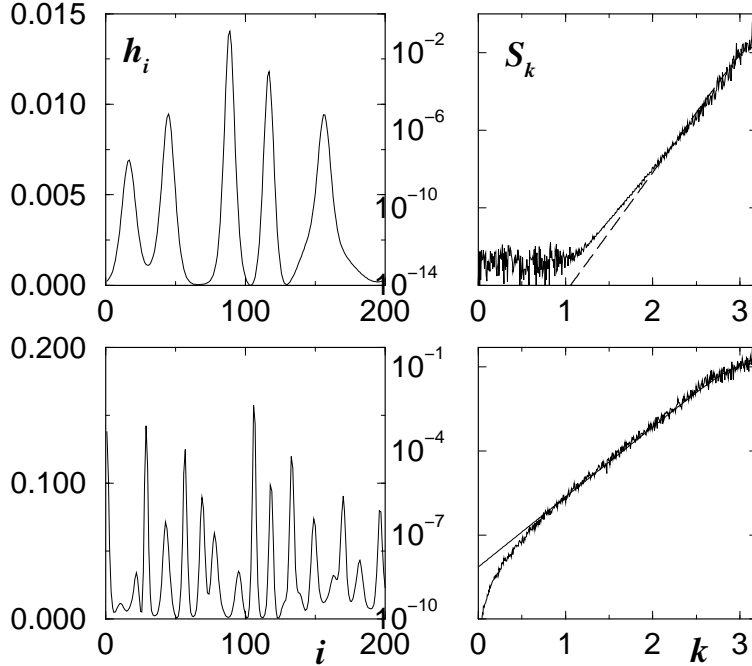,width=10cm}
\caption{
The soliton gas generated after the instability for $N=1024$ 
for  $\varepsilon=0.0033$, $t=8000$ (upper panels)
and $\varepsilon=0.0330$, $t=800$ (lower panels).
The corresponding power spectra $S_k$ of the particles' velocities 
(right panels) averaged over a time interval $\approx 100$. The
dashed lines are the best fit with the formula (\protect\ref{sksol}).
}
\label{esoli}
\end{figure}

As already mentioned, the modulational instability produces a strong 
localization of the vibrational energy along the chain (see  Fig. \ref{esoli}).
We wish now to describe such a set of localized objects. More precisely we 
will show that it is basically a gas of ESs and study some of its properties.

Let us consider a set of $N_{s}$ non-overlapping ESs of the 
form (\ref{soliton}) centered at the points $x_{l}$ along the line 
and with initial phases $\varphi_{l}$. For the sake of simplicity, let us
also assume them to have approximatively the same amplitude. According to 
(\ref{spesol}), the spatial Fourier transform of the corresponding displacement 
field will be  
\begin{equation}
u(k,t)=\frac{1}{2}
\sum_{l=1}^{N_{s}}\cos(\omega_{s} t +\varphi_{l})
\left[F_s(k-\pi,x_l) + F_s(k+\pi,x_l) \right]~.
\end{equation} 
The velocity power spectrum $S_{k}$ as defined by Eq. (\ref{sk}) in the 
discrete case, can thus be approximated as (we use the explicit form of 
$F_s$)   
\begin{eqnarray}
S_{k}\;=\;\frac{1}{N}\langle\mid \dot{u}(k,t)\mid^2\rangle 
&=& \frac{\pi^2\omega_{s}^2}{48N} \{
\frac{1}{\cosh^{2}[\frac{\pi(\pi-k)}{\phi_{max}2\sqrt{6}}]}
\mid\sum_{l}
e^{i\varphi_{l}+i(k-\pi)x_{l}}\mid^{2} \nonumber \\
&&+\frac{1}{\cosh^{2}[\frac{\pi(k+\pi)}{\phi_{max}2\sqrt{6}}]}
\mid\sum_{l}
e^{i\varphi_{l}-i(k+\pi)x_{l}}\mid^{2}\}~,
\label{powersol}
\end{eqnarray}       
where the angle brackets denote the averaging over a time interval larger
than $2\pi/\omega_s$ (that is performed to remove the oscillations as well as
statistical fluctuations).  In the last expression we neglected  
exponentially small overlap between the components of (\ref{spesol})   
localized respectively around $k=\pi$ and $k=-\pi$.  

Finding a more explicit expression for $S_k$ requires of
course some hypotesis on the statistical properties of the  positions
$x_{l}$  and phases $\varphi_{l}$. For instance, in the limit of completely 
uncorrelated ESs one can replace the square modulus of each sum in 
Eq. (\ref{powersol}) by $N_s$. To take into 
account weak partial correlations, we introduce a 
phenomenological parameter $p$ ($p\sim 1$) and replace the mentioned 
square modulus by $pN_s$. Obviously, $p=1$ corresponds to a completely 
random case while the cases $p>1$ or $p<1$  describe, respectively, 
the arrays of the partially correlated or anticorrelated ESs.  
With such a definition 
\begin{equation}
S_{k}=pn_{s}\frac{\pi^2\omega_{s}^2}{48}\{
\frac{1}{\cosh^{2}[\frac{\pi(\pi-k)}{\phi_{max}2\sqrt{6}}]}
+\frac{1}{\cosh^{2}[\frac{\pi(k+\pi)}{\phi_{max}2\sqrt{6}}]}\}~,
\label{specsol}
\end{equation}          
where $n_{s}=N_{s}/N$ is a density of the envelope solitons. 

To compare the latter expression with the numerically measured  
$S_k$, one has to relate $\omega_s$, $n_{s}$ and $\phi_{max}$  in 
Eq. (\ref{specsol}) with the energy density  
$\epsilon$. As we deal with the limit $\epsilon\ll 1$, one has that 
$\omega_s\approx 2$ since $A\approx\sqrt{2\epsilon}/2$ (cf. Eq.(\ref{ic}) and 
the definition of the amplitude $A$). Moreover, the density $n_{s}$ 
will be given by the inverse wavelength of the modulational instability, 
namely
\begin{equation}
n_{s}=\frac{k_{m}^{\star}}{2\pi}=\frac{1}{\pi}\sqrt{\frac{3\epsilon}{2}}~.
\label{ns}
\end{equation}
This amounts to say that a single localized object emerges from each 
wavelength. A similar reasoning allows us to relate the total energy 
$E_{s}$ (which is basically proportional to $\phi_{max}$, see again 
Eq. (\ref{es2})) to $\epsilon$ as
\begin{equation}
E_{s}=q\frac{\epsilon}{n_{s}}=
q\, 2\pi\sqrt{\frac{\epsilon}{6}}~.
\label{es}
\end{equation}    
The phenomenological parameter $q$, $0\leq q\leq 1$, 
gives the fraction of the initial vibrational energy which is absorbed by 
the set of ESs (and actually accounts for the partial overlap between them). 
In particular, for $\epsilon\to 0$ we expect the soliton gas to be more and 
more rarified and to contain almost all of the initial energy in the lattice 
so that $q\to 1$. 

A consequence of Eqs. (\ref{ns}) and (\ref{es}) is that the ratio 
between the total energy of the ESs and their density is (almost) independent 
on $\epsilon$: 
\begin{equation}
\frac{E_{s}}{n_{s}}=q\frac{2\pi^2}{3}~.
\label{ratio}
\end{equation} 

\begin{figure} 
\centering\epsfig{figure=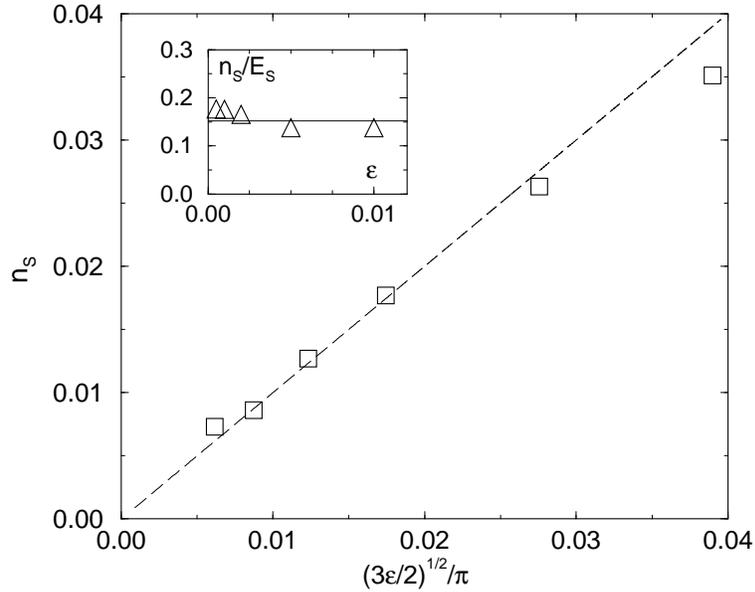,width=10cm}
\caption{
The maximal initial density $n_s$ of ESs as a function of the
energy density for $N=1024$-. 
The solid line corresponds to formula (\protect\ref{ns}).
The inset shows the ratio between the number and energy of
the solitons: the horizontal line corresponds to Eq. (\protect\ref{ratio})
with $q=1$.
}
\label{nrs}
\end{figure}

Before proceeding further, we numerically checked that the above picture is 
correct by evaluating the number and energy of solitons as a function of the 
initial energy density. The results reported in Fig. \ref{nrs} show a very good
agreement with both Eqs. (\ref{ns}) and (\ref{ratio}). Notice that, as expected,
the data are very well fitted with $q=1$ at least for $\epsilon<0.01$.

Finally, using Eqs. (\ref{specsol}), (\ref{ns}), (\ref{es}), and (\ref{es2}), 
we obtain the desired expression for the velocity power spectrum as a function of 
$\epsilon$ and the (unknown) parameters $p$ and $q$ only:
\begin{equation}
S_{k}=p\frac{\pi}{16}\sqrt{\frac{\epsilon}{6}}\omega_{s}^2
\{\frac{1}{\cosh^2
[\frac{\pi-k}{q\sqrt{6\epsilon}}]}+\frac{1}{\cosh^2
[\frac{\pi+k}{q\sqrt{6\epsilon}}]}\}~,
\label{sksol}
\end{equation}
where $\omega_{s}=2$.

We obtained  a very good fit of our numerical data, for energy density 
$\epsilon=0.0033$, using Eq. (\ref{sksol}) with $p=0.50$ and $q=1.0$ 
(see the right panels of Fig. \ref{esoli}). We interpret this by saying that 
the system is assimilated to a diluted gas of nonoverlapping and partially 
anticorrelated (i.e. out of phase) ESs. For larger initial energy density 
$\epsilon=0.033$ the data are fitted by $p=1.12$ and $q=0.78$ (lower panels of 
Fig. \ref{esoli}). 
This latter case correspond, as expected, to a more dense gas of partially 
overlapping ESs, which is characterized, in comparison with the previous one,  
mainly by pair correlations between neighbours. 

\section{The second localization stage}

To understand why a set of almost nonoverlapping small-amplitude ES finally 
merge into a large-amplitude one, it is first of all convenient to recast the 
second order (in time) equation (\ref{envelope}) for the real envelope-function 
in two first order equations. For small-amplitude oscillations with a 
frequency close to the band-edge $\omega_{max}=2$ this is 
accomplished introducing the complex function $\psi(x,t)$ such that
\begin{equation}
f(x,t)=\frac{1}{2}[\psi(x,t)e^{-i\omega_{max}t}\,+\,\psi^{\ast}(x,t)
e^{i\omega_{max}t}]~. \nonumber
\end{equation}
Substituting this expression into Eq. (\ref{envelope}) and assuming that 
$\dot\psi\ll\omega_{max}\psi$, after some algebra one can
obtain (in the RWA) the following coupled nonlinear-Schr\"odinger-type 
equations for $\psi$ and $\psi^{\ast}$ 
\begin{eqnarray}
&&-i\omega_{max}\dot\psi+\frac{1}{2}\psi_{xx}+
\frac{1}{24}\psi_{xxxx}+6\mid\psi\mid^2\psi \nonumber \\
&&+
\frac{3}{2}[2\psi(\psi^{\ast}\psi_{xx}+\mid\psi_{x}\mid^2)+
\psi^2\psi^{\ast}_{xx}+\psi^{\ast}\psi_{x}^2]=0~, \nonumber \\
&& i\omega_{max}\dot\psi^{\ast}+\frac{1}{2}\psi^{\ast}_{xx}+
\frac{1}{24}\psi^{\ast}_{xxxx}+6\mid\psi\mid^2\psi^{\ast}
\nonumber \\
&&+
\frac{3}{2}[2\psi^{\ast}(\psi\psi_{xx}^{\ast}+\mid\psi_{x}\mid^2)+
\psi^{\ast 2}\psi_{xx}+\psi\psi_{x}^{\ast 2}]=0~.
\label{schroed}
\end{eqnarray}  
Equations (\ref{schroed}) can be expressed in canonical form as
$$
i\omega_{max}\dot\psi=\frac{\delta{\cal H}}{\delta\psi^{\ast}},~~~
- i\omega_{max}\dot\psi^{\ast}=\frac{\delta{\cal H}}{\delta\psi}~,
$$
where the Hamiltonian ${\cal H}=\int Hdx$ has the following density $H$: 
\begin{eqnarray}
H&=&-\frac{1}{2}[\mid\psi_{x}\mid^2-\frac{1}{12}
\mid\psi_{xx}\mid^2-6\mid\psi\mid^4 \nonumber \\
&&+6[\mid\psi\mid^2\mid\psi_{x}\mid^2+
\frac{1}{4}(\psi^2\psi_{x}^{\ast 2}+\psi_{x}^2\psi^{\ast 2})]]~.
\label{hamschro} 
\end{eqnarray} 
One can ascertain that equations (\ref{schroed}) admit the  
integrals of motion
\begin{equation}
{\cal H}=\int Hdx~, ~~~{\cal P}=-\frac{i}{2}\int [\psi\psi_{x}^{\ast}-\psi^{\ast}
\psi_{x}]dx~, ~~~{\cal N}=\int \mid\psi\mid^2dx~,
\label{quanta}
\end{equation}
which are the energy, momentum and number of quanta, respectively. 

The inclusion of higher-order and nonlinear dispersive terms make 
equations (\ref{schroed}) nonintegrable 
(i.e. they  do not have an infinite number of constants of motion). This 
simple fact, together with the conservation laws (\ref{quanta}) has deep 
consequences on the dynamics: it is in fact known \cite{ZakhYank} that in this 
case the total number of solitons is not conserved. 
Actually, they can merge together enhancing their amplitudes and decreasing 
their number due to an exchange of both energy and number of quanta. Indeed, 
the Hamiltonian (\ref{hamschro}) describes an effective $repulsive$ 
interaction between quasiparticles with $negative$ (and amplitude-dependent) 
effective mass and it is therefore equivalent to a system of quasiparticles 
with $positive$ effective mass and effective $attractive$ interaction. 
As the solitons merge, the released binding energy is carried away by free 
waves and the ``phononic'' part of the spectrum is gradually populated.

More precisely, the kinematics of the soliton-soliton interaction process
requires an increase  of the number of quanta in the stronger soliton and 
a decrease in the weaker one accompanied by the release of some binding 
energy by means of free plane waves that give also the recoil momentum to 
the solitons \cite{ZakhYank}.  Accordingly, the reverse process should 
include a triple collision (two solitons and a free wave) and
therefore is much less probable than the direct one. It is therefore 
clear that eventually only one large-amplitude  soliton survives on the 
background of small-amplitude free waves with a continuous frequency 
spectrum. This argument explains qualitatively why the
purely dynamical process of soliton merging is time-irreversible.

The above reasoning can be further carried on to give a more quantitative 
description of the phenomenology. For instance, it is possible to explain
the observed fact that the characteristic time for complete soliton merging
in the quartic FPU chain scales with the energy density as $\epsilon^{-2}$
\cite{Stefano} (see also \cite{Burlakov}). To this aim, let us estimate the ES lifetime (or relaxation
time) $\tau_{ss}$ due to the scattering mentioned above. It can be written
as $\tau_{ss}= l/\bar{v}_{s}$, where $l$ and $\bar{v}_{s}$ 
are respectively the mean free path and average velocity of the ES during 
the relaxation process. The mean free path will be in turn given by 
$l\sim 1/\sigma_{ss}n_{s}$, where $n_{s}$ is a density of ESs and $\sigma_{ss}$
is the scattering cross-section.

If we assume the lifetime to be much larger than 
the inverse frequency of the soliton ($\tau_{ss}\gg 1/\omega_{s}$), we can 
resort to the Born approximation to estimate $\sigma_{ss}$. As time goes by
and the soliton gas rarifies due to the process of merging, the approximation 
becomes more and more justified . Within this perturbative 
treatment and in the low velocity limit, we thus find that
$\sigma_{ss}$ is simply proportional to the square modulus of $\int U_{int}dx$, 
where $U_{int}$ is the interaction potential. According to (\ref{hamidens}), 
the latter is proportional to the product of the envelope functions
of two small-amplitude ESs: $U_{int}\propto f_{1}f_{2}$. Assuming for simplicity
that both of them have approximatively the same amplitude $\phi_{max}$, it is easy 
to ascertain from (\ref{soliton}) that the overlap integral scales as 
$\phi_{max}$, similar to the energy of a single small-amplitude ES
(\ref{es2}). 
Therefore, the
cross-section $\sigma_{ss}$ is proportional to $\phi_{max}^2$ and, according to
Eqs. (\ref{es}) and (\ref{es2}), this implies that $\sigma_{ss}\propto\epsilon$.  
Finally, as $n_{s}\propto\sqrt{\epsilon}$, see Eq. (\ref{ns}), we obtain that
$l\propto\epsilon^{-1.5}$.  

To estimate $\bar{v}_{s}$, we compute first the energy of the moving soliton of 
Eqs. (\ref{schroed}). In the small-amplitude limit, when one can neglect the 
higher-order and nonlinear dispersive terms, Eqs. (\ref{schroed}) admit the 
exact solution \cite{KosKov,ZakhYank}
\begin{equation}
\psi(x,t)=\psi_{max}\frac{\exp[i(kx-\Omega
t)]}{\cosh[(x-V_{g}t)\psi_{max}\sqrt{6}]},
\label{solpsi}
\end{equation}
where $V_{g}$ is a group velocity of the soliton, $k=-V_{g}\omega_{max}$ is a 
soliton quasimomentum and $\Omega=(3\psi_{max}^2-k^2/2)/\omega_{max}$. Using 
Eqs. (\ref{hamschro}), (\ref{quanta}),  and (\ref{solpsi}), we find 
the soliton number of quanta ${\cal N}_{s}$, momentum ${\cal P}_{s}$, and 
energy ${\cal E}_{s}$ (for $\omega_{max}=2$):
\begin{equation}
{\cal N}_{s}=\psi_{max}\sqrt{\frac{2}{3}}, ~~{\cal P}_{s}=-k{\cal N}_{s}
=2V_{g}{\cal N}_{s},
~~{\cal E}_{s}=\frac{3}{2}{\cal N}_{s}^3-2{\cal N}_{s}V_{g}^2=\frac{3}{2}{\cal
N}_{s}^3-\frac{{\cal P}_{s}^2}{2{\cal N}_{s}}.
\label{squanta}
\end{equation}
These relations show that small-amplitude moving soliton can be indeed 
considered as a bound state of ${\cal N}_{s}$ negative-effective-mass 
quasiparticles (harmonic modes) with a correspondingly 
$positive$ binding energy equal to $\frac{3}{2}{\cal N}_{s}^3$ 
(cf. \cite{KosKov1}). 
  
If we compare the definition of ${\cal N}_{s}$ in (\ref{squanta}) with 
expressions (\ref{es}) and (\ref{es2}) for the energy of the ES (\ref{soliton}),
we conclude that ${\cal N}_{s}\propto\sqrt{\epsilon}$. Let us now assume that a 
kind of virial theorem holds for the set of solitons moving along the chain with 
periodic (or fixed) boundary conditions. This amounts to say that (time) averaged 
kinetic and potential energies of such finite dynamical system should be proportional 
(see, e.g., $\cite{L.L.Mech}$). It then follows from the definition 
(\ref{squanta}) of ${\cal E}_{s}$ that 
$\langle V_{g}^2\rangle\sim\langle{\cal N}_{s}^2\rangle\propto\epsilon$. 
The validity of such a relation can be physically justified by observing that,  
as a result of the exchange of the number of quanta during the scattering 
processes, the solitons receive the recoil quasimomentum $k$ from free waves 
which in turn changes the soliton velocities thus providing some form of 
``thermalization''. From all the above follows that 
$\bar{v}_{s}\sim\sqrt{\langle{V}_{g}^2\rangle}\propto\sqrt{\epsilon}$. 

Finally, we obtain the estimate $\tau_{ss}=l/\bar{v}_{s}\propto\epsilon^{-2}$ 
and correspondingly  $\tau_{ss}\gg 1/\omega_{s}\approx 1/2$, that is fully 
consistent with the initial assumptions. As already mentioned, this scaling
is in agreement with the numerical observation $\cite{Burlakov,Stefano}$.

\section{The breather}

\begin{figure} 
\centering\epsfig{figure=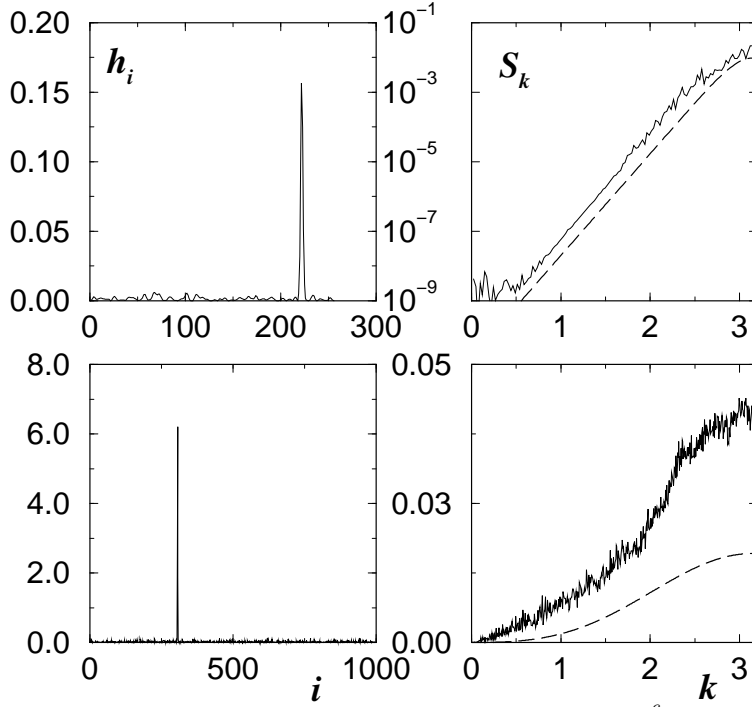,width=10cm}
\caption{
Final breather for different energies  $\varepsilon=0.0033$, $N=256$, 
$t=1.2 \, 10^6$ (upper panels) and $\epsilon=0.0330$, $N=1024$, $t=8\, 10^4$ 
(lower panels). The last spectrum has been averaged over a time $10^4$.
Notice the different vertical scales. Dashed lines correspond to Eqs.
(\protect\ref{spesol}), (\protect\ref{spesol2}) and (\protect\ref{spebreaex}) with the numerically
measured value of $\phi_{max}$ and $k_b=\pi/3$.
}
\label{brea}
\end{figure}

\begin{figure} 
\centering\epsfig{figure=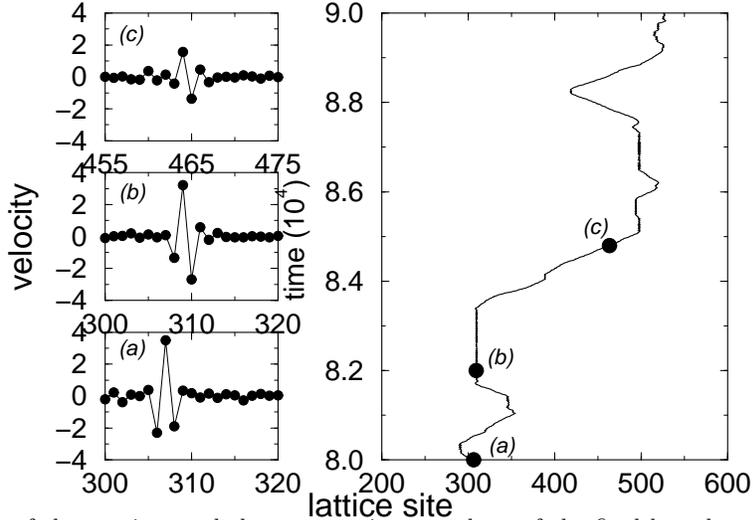,width=10cm}
\caption{
The position of the maxima and three successive snapshots of the final breather 
as a function of time for $\epsilon=0.0330$ and $N=1024$. During this observation 
time the value of the breather energy is practically constant.
}
\label{position}
\end{figure}

In the previous Section we explained why the scattering favors the creation
of a single localized object. We now give a brief account of some of its 
properties. The $h_n$ patterns reported in the left panels of 
Figs. \ref{brea} and \ref{position} illustrate the remarkable fact that a 
macroscopic fraction
of the total energy accumulates on a few sites. For instance for the case
$\epsilon=0.033$ and $N=1024$, only $three$ sites bear an energy 10.9 corresponding to about 
30\% of the total. As expected from the previous Section, the remaining is 
distributed along the chain in the form of small amplitude (harmonic) oscillations
(see again Fig. \ref{brea})

If we observe the breather on a time scale during which its amplitude
does not change considerably, we see that it randomly alternates
between a standing and moving state. Fig. \ref{position} clearly shows  
how each motion occurs with an almost constant velocity and is accompanied
by changes of the shape. For a given amplitude, the modulus of such a 
velocity is always the same. We checked that its value is 
consistent with previous studies on exact breather solution:
for example from Fig. \ref{position} we found indeed the velocity to be
0.12(8) in excellent agreement with the value 0.120 obtained interpolating
the data of Ref. \cite{Sand.Page}.

The main issue concerning the velocity is the one of its selection. Although
we will not enter into details here, we just anticipate that some analytic clue
can be achieved by studying the moving solutions of Eq. (\ref{envelope})
\cite{KosPRB}. In particular, we expect the velocity to increase 
with the amplitude (almost linearly in the large-amplitude limit 
$9\phi_{max}^2\gg 1$), a fact that is actually 
consistent with our as well as with other simulations \cite{Stefano,Bourbon}.

Let us now discuss the power spectra of those strongly localized states. 
The right panels of Fig. \ref{brea} show how they differ depending on the
amplitude. At smaller energies (upper panels) they resemble more closely
an ES and the spectrum would have the exponentially decaying prescribed
by (\ref{spesol}). Upon increasing the energy, $S_k$ acquires a  
sinusoidal component like (\ref{spebreaex}) characteristic of a breather 
solution. This is simply the consequence of the fact that in the continuum
approach the two are the only limiting cases of the general solution (\ref{implicit})
with the amplitude $\phi_{max}$ being determined only by the total 
energy of the lattice.

Obviously, the existence of the vibrational 
background affects the structure of the
spectra. 
As the background contains a non-negligible energy (the remaining 70\% in the
case mentioned above), a complete description of the spectra requires some 
information on its features. One consequence that we can draw from  Fig. \ref{brea} 
is that the naive picture of a single breather moving through a gas of small waves 
is oversimplified. In fact this would imply the spectrum to be a
sum of a part like (\ref{spebrea}) plus an almost flat part. The form of $S_k$ seems
rather to indicate that only the high-frequency domain of the harmonic spectrum is 
populated. 

To give a complete description of the relaxation process, we should now consider
the final stage of the breather destruction (see also Ref. \cite{Stefano} for 
a discussion). This requires understanding how energy 
is transferred from large to small wavenumbers. This issue involves in turn the 
interaction of a large-amplitude ES with small-amplitude plane waves. The latter 
problem has been studied in Ref.
\cite{cretegny}. Here we limit ourselves to observe that it is still not 
clear how to explain why the tipical time scale for such a process should 
scale again as $\epsilon^{-2}$. An analytic argument for that can be actually 
given
only in the very last stage, when the system is sufficiently close to equilibrium
and relaxation of Fourier modes can be described by usual linear-response 
theory \cite{lepri}.
 
\section{Finite-temperature effects}

What has been sofar discussed refers to a situation where all the
lattice energy is initially fed in the boundary mode. This is clearly 
not realistic for actual experimental conditions, when the system is
in contact with a thermal bath which populates all the other modes
even at (classically) low temperatures. One may therefore wonder to what extent
the mechanisms described above are effective in localizing the energy.

In the present section we try to approach the problem by first considering 
a more general class of initial conditions defined by replacing 
second expression in (\ref{ic}) by 
\begin{equation}
\dot u_{n}(0) \;=\; (-1)^{n} V +\sigma'\xi_{n}~,
\label{velocity}
\end{equation}
where $\xi_{l}$ is a gaussian random number such that $\langle\xi_{l}\rangle=0$ and  
$\langle\xi_{l}\xi_{l+m}\rangle=\delta_{m0}$. In order to assign the  
initial energy per particle to be equal to $\epsilon$, we choose 
$V=2\epsilon/\sqrt{2\epsilon+\sigma^2}$ and $\sigma^{'}=
\sigma\sqrt{2\epsilon/(2\epsilon+\sigma^2)}$. 
The parameter $\sigma$ controls thus the ``level of noise" in the initial 
conditions. Indeed, the limit $\sigma^2\ll\epsilon$
(exactly the one considered in the previous simulations) amounts to 
feeding all the available energy in the Brillouin-zone-boundary mode 
whereas upon increasing  $\sigma$, we can initially excite also the other 
Fourier modes of the chain. With our definition, equilibrium
(i.e. equipartition of energy) is attained for $\sigma\to\infty$ (actually 
when $\sigma^2\rightarrow\sigma_{max}^2=2\epsilon N\gg\epsilon$). 

\begin{figure} 
\centering\epsfig{figure=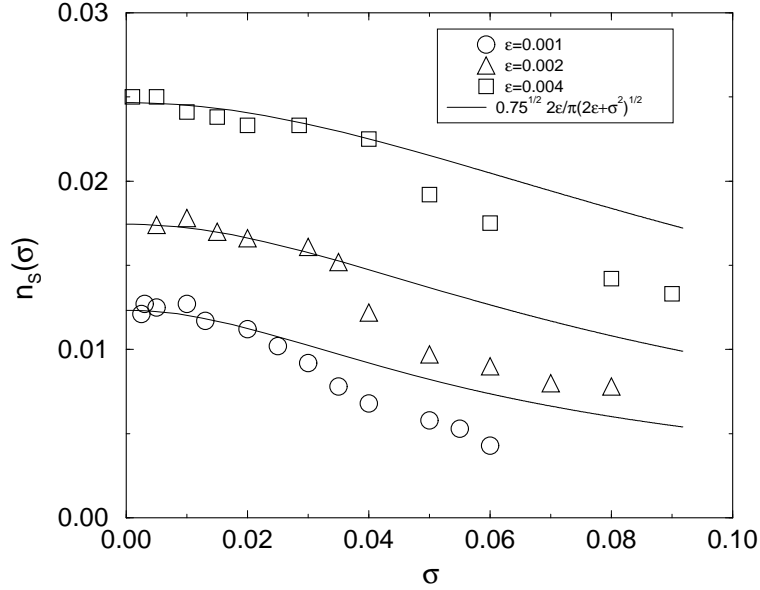,width=10cm}
\caption{
The maximal initial density of ESs $n_S$ as a function of the 
parameter $\sigma$ for several values of $N$ ranging between 920 and 1024.
Solid lines correspond to formula (\protect\ref{nss}).
}
\label{nssigma}
\end{figure}

The results presented before can be first extended to the finite $\sigma$- 
values by the following reasoning. We showed that the initial density of ESs
is basically determined by the vibrational energy available in the
Brillouin-zone-boundary mode. Therefore, one can straightforwardly replace 
$\sqrt{2\epsilon}$ by $V$ in Eq. (\ref{ns}) obtaining
\begin{equation}
n_{s}(\epsilon,\sigma)=\frac{1}{\pi}\sqrt{\frac{3}{4}} \, 
{2\epsilon\over\sqrt{2\epsilon+\sigma^2}}\quad .
\label{nss}
\end{equation}  
As seen in Fig. \ref{nssigma}, the numerical data are in good agreement with 
this formula, at least for moderate amplitudes $\sigma^2\lesssim\epsilon/2$. 
This indicates that the mechanism of vibrational energy localization, induced 
by the modulational instability is relatively robust, at least with respect 
to those type of perturbation in the initial conditions. 

Remarkably, as soon as $\sigma$ overcomes the value $\epsilon/2$, 
$n_{s}$ seems to be slightly less (around 10 \%) than what is 
expected from the purely energetic argument leading to (\ref{nss}). 
Although in this intermediate regime the numerical measurements are 
definitely less accurate due to the large vibrational background, 
we can at least give a qualitative justification to this.
In fact, in this case we expect some effective dissipative 
mechanisms to enter into play and some of the localized excitation will 
be effectively damped out by the interaction with the ``thermal bath'' 
of the other vibrational modes \cite{lepri}.

\begin{figure} 
\centering\epsfig{figure=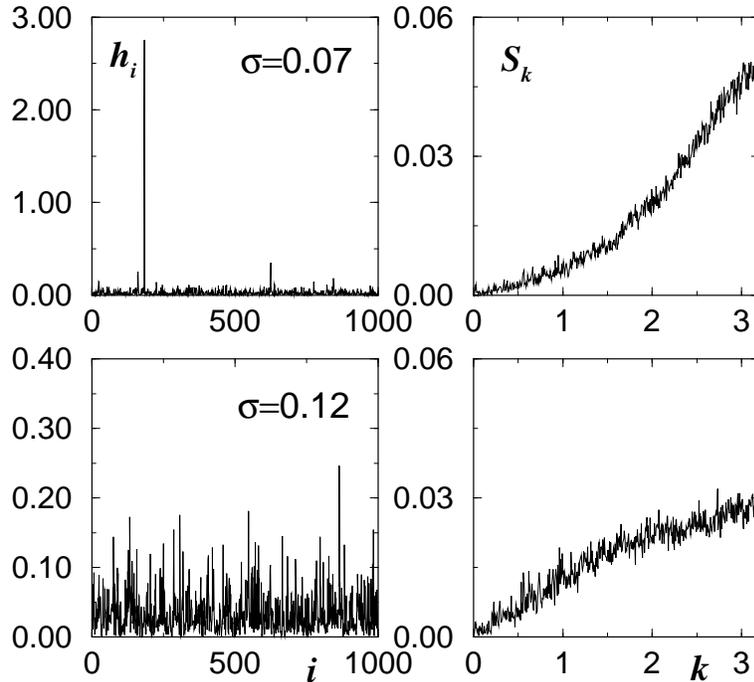,width=10cm}
\caption{
The distribution of energy densities and power spectra of the velocities for 
$\epsilon=0.0330$, $N=1024$, $t=8\, 10^4$ and two different amplitudes of
initial perturbation $\sigma = 0.07$ (upper panels) and 0.12(lower panels). 
}
\label{sigbrea}
\end{figure}

In order to better clarify this, we performed also some measurements of 
the power spectrum (\ref{sk}) for the relatively large values of $\sigma$ 
(see Fig. \ref{sigbrea}). The results confirm that even
for $\sigma^2\approx\epsilon/2$ (lower panels) some kind of 
``localization" is still present, although it is hardly detectable by 
direct measurements. 

To summarize, the creation 
of localized excitations, induced by the modulational instability, 
is more and more inhibited the closer one starts to the equipartition (thermal
equilibrium) state. This is caused both by the less amount of energy 
available in the corresponding short-wavelength mode (as $V\rightarrow 0$ 
for $\sigma^2\gg\epsilon$, see Eq. (\ref{velocity})) 
and by the effective dissipation caused by the interaction of small-amplitude 
ESs with the vibrational background. At least in principle, the 
envelope-function approach could be a useful starting point for a 
more detailed analysis once one would be able to take into account the 
fluctuating background by casting it in suitable stochastic terms.

\acknowledgments
 The authors are grateful to I. Barashenkov, S. Flach, A.S. Kovalev, A. Mayer,
A. Politi and S. Ruffo for the useful
discussions. This work was supported by the Max Planck Society.

\newpage 

\end{document}